\documentclass[a4paper,twocolumn]{article}

\title{Exhaustive Search for Low Autocorrelation Binary Sequences}

\author{S.~Mertens\thanks{email: stephan.mertens@physik.uni-magdeburg.de}\\[1ex]
        \small Institut f\"ur Theoretische Physik\\
				\small Otto-von-Guericke Universit\"at, Postfach 4120, D-39016 Magdeburg, Germany
			 }
			 
\usepackage{graphicx,fancyheadings,algorithm,algorithmic,amssymb}
\date{June 1996}

\floatstyle{ruled}
\restylefloat{figure}
\restylefloat{algorithm}
\restylefloat{table}

\begin{document}

\maketitle

{\bf Abstract: }{
Binary sequences with low autocorrelations are important in communication
engineering and in statistical mechanics as groundstates of the Bernasconi-model. 
Computer searches
are the main tool to construct such sequences. Due to the exponential size $O(2^N)$
of the configuration space, exhaustive searches are limited to
short sequences.
We discuss an exhaustive search algorithm with run time characteristic
$O(1.85^N)$ and apply it to compile a table of exact groundstates of the
Bernasconi-model up to $N=48$. The data suggests $F> 9$ for the optimal merit
factor in the limit $N\to\infty$.
}

\section{Introduction}
Binary sequences $S = \{s_1\!=\!\pm1,\ldots,s_N\}$ with low off-peak
autocorrelations
\begin{equation}
\label{open_correlations}
C_k(S) = \sum_{i=1}^{N-k}s_is_{i+k}.
\end{equation}
have applications in many
communication engineering problems \cite{schroeder:84}. One exciting example
has been their use in high precision interplanetary
radar measurements to check out space-time-curvature \cite{shapiro:etal:68}.

Physicists prefer to consider binary sequences as one dimensional systems of
Ising-spins. In this context,
low autocorrelation binary sequences appear as minima of the
energy 
\begin{equation}
\label{bernasconi}
E(S) = \sum_{k=1}^{N-1} C_k^2(S).
\end{equation}
This is the Bernasconi-model \cite{bernasconi:87}.
It has long-range 4-spin interactions and is completely
deterministic, i.e. there is no explicit or
quenched disorder like in spin-glasses. Ne\-ver\-the\-less the ground states are
highly disordered -- quasi by definition.
This self-induced disorder resembles very much the situation in real glasses.
In fact, the Bernasconi-model exhibits features of a glass transition like a
jump in the specific heat \cite{bernasconi:87} and slow dynamics and aging
\cite{krauth:mezard:95}.

A clever variation of the replica method allows an ana\-ly\-ti\-cal treatment of the
Bernasconi-model in the high-temperature regime \cite{bouchaud:mezard:94,marinari:parisi:ritort:94a}.
For the low-temperature re\-gime, analytical results are rare -- especially
the ground states are not known.
With periodic boundary conditions, i.e.\ with
\begin{equation}
\label{periodic_correlations}
C_k = \sum_{i=1}^{N}s_is_{(i+k-1)(\bmod N) + 1},
\end{equation}
instead of eq.~\ref{open_correlations},
the construction of ground states is possible for special values of $N$.
Example: For $N=4n+3$ prime, the modified Legendre-sequence
\begin{equation}
\label{legendre}
s_j = \left\{
\begin{array}{cl}
 j^{\frac12(N-1)}\bmod N & 1 \leq i < N\\
\pm1 & i = N
\end{array}
\right.
\end{equation}
yields $C_k^2 = 1$, the minimum possible value for odd $N$.
Other ground states can be constructed from linear shift register sequences
based on primitive polynomials over Galois fields. This construction requires $N = 2^p - 1$
with $p$ prime. See \cite{schroeder:84,marinari:parisi:ritort:94a} for details.

For the model with open boundary conditions (eq.~\ref{open_correlations})
no construction of ground states is known, not even for special values of $N$.
The Legendre-sequences are far from the true ground states \cite{golay:83}.
The only exact results have been provided by exhaustive enumerations, limited
however by the exponential complexity of the problem to systems 
smaller than $N = 32$ \cite{golay:82}.
Partial enumerations allow larger values of $N$ but cannot
guarantee to yield true ground states.
Promising candidates for partial enumerations are the skew-symmetric sequences
of odd length $N = 2n-1$. These sequences satisfy
\begin{equation}
\label{skew-symmetry}
s_{n+l} = (-1)^l s_{n-l}, \quad l = 1,\ldots,n-1
\end{equation}
from which it follows that all $C_k$ with $k$ odd vanish. The restriction to skew-symmetric
sequences reduces the effective size of the problem by a factor of 2, but the true
ground states are {\em not} skew-symmetric for several values of $N$, as we will see below.

Finding the ground states of the Bernasconi-model has turned out to be a hard
mathematical problem. Golay \cite{golay:82,bernasconi:87} has conjectured that the maximal
merit factor
\begin{equation}
\label{merit}
F = \frac{N^2}{2E}
\end{equation}
should obey $F\lesssim 12.32$ for $N\gg 1$. However,
heuristic
searches among skew-symmetric sequences up to $N = 199$ suggest
$F\approx 6$ for long sequences \cite{beenker:claasen:hermens:85}, a value consistent with results
from simulated annealing \cite{bernasconi:87}. This large discrepancy indicates
that the groundstates, i.e.\ the sequences with high merit factors
$6 < F \lesssim 12$, must be extremely isolated
energy minima in configuration space. Stochastic search procedures including
simulated annealing are not well suited to find these
``golf-holes''. Exhaustive search seems to be the only approach
at least for small systems. The complete configuration space
has been searched up to $N=32$ \cite{golay:82}, the skew-symmetric subspace
up to $N=71$ \cite{golay:77,degroot:etal:92}. Fifty days of CPU-time on a
special purpose computer have been used for an exhaustive search for binary
sequences up to $N=40$ that minimize $\max_{k}|C_k|$ \cite{lindner:75}.

In this contribution we discuss a fast algorithm for the exhaustive enumeration.
It is fast enough to
yield exact groundstates of the Bernasconi-model up to $N=48$ and can easily
be modified for partial enumerations.
The data is used to estimate the optimal
merit factor in the large $N$ limit. 

\section{The Algorithm}

Any algorithm that performs an exhaustive search for the ground state of the Bernasconi-model
has to cope with the enormous size ($2^N$) of the configuration space. This exponential complexity
limits the accessible values of $N$ very soon and calls for methods to restrict the search
to smaller subspaces without missing the true ground state. {\em Symmetries} are an abvious device
to cut out portions of the configuration space. We will see below, that the use
of symmetries can reduce the size of the search space by a factor of about $1/8$.
A method borrowed from combinatorial optimization - {\em branch and bound} - will prove
useful to reduce the complexity from $O(2^N)$ to $O(b^N)$ with $b < 2$. We will further see,
that the enumeration problem is suited almost perfectly for {\em
parallelization}.

\subsection{Symmetries}

The correlations $C_k$ (eq.~\ref{open_correlations}) are unchanged when the sequence is
complemented or reversed. When alternate elements of the sequence are complemented, the
even-indexed correlations are not affected, the odd-indexed correlations only change sign.
Hence, with the exception of a small number of symmetric sequences, the $2^N$ sequences
will come in classes of eight which are equivalent. The total number of nonequivalent
sequences is slightly larger than $2^{N-3}$.

The $m$ left- and $m$ rightmost elements of the sequence can be used to parameterize the 
symmetry-classes. For $m = 3$ and $N$ odd, this gives 12 classes:
\begin{displaymath}
\begin{array}{cc}
---\cdots--- &\quad ---\cdots++- \\
---\cdots--+ &\quad ---\cdots+++ \\
---\cdots-+- &\quad --+\cdots--+ \\
---\cdots-++ &\quad --+\cdots-++ \\
---\cdots+-- &\quad --+\cdots+-- \\
---\cdots+-+ &\quad --+\cdots++- 
\end{array}
\end{displaymath}
For $N$ even there are 10 classes. In general the number $c$ of symmetry-classes that can be
distingiushed by $m$ left- and $m$ right-border elements reads
\begin{equation}
   \label{classes}
   c(m) = 2^{2m-3} + 2^{m-2+(N\bmod2)}
\end{equation}
and the number of nonequivalent configurations reduces to a fraction
\begin{equation}
   \label{fraction}
   \frac{c(m)}{2^{2m}} = \frac18 + \frac1{2^{m+2-(N\bmod2)}}.
\end{equation}
The optimal value $\frac18$ is approached with increasing $m$.

\subsection{Branch and Bound}

Branch and bound methods are commonly used in combinatorial optimization
\cite{balas:toth:85} and (less frequently) in statistical mechanics
\cite{kobe:hartwig:78,hartwig:daske:kobe:84}.
They solve a discrete optimization problem by breaking up its feasible
set into successively smaller subsets ({\em branch}), calculating bounds on the
objective function value over each subset, and using them to discard certain
subsets from further consideration ({\em bound}).
The procedure ends when each subset has either produced a feasible solution,
or has been shown to contain no better solution than the one already in hand.
The best solution found during this procedure is a global optimum.

The idea is of course to discard many subsets as early as possible during the
branching process, i.e.\ to discard most of the feasible solutions before
actually evaluating them. The success of this approach depends on the 
branching rule and very much on the quality of the bounds, but it can
be quite dramatic. Numerical investigations have shown, for example, that
the  $n$-city Traveling Salesman Problem can be solved exactly
in time $O(n^\alpha)$ with $\alpha < 3$ using branch and bound methods
\cite{balas:toth:85}! This is no contradiction to the exponential complexity of the TSP
since the latter is the guaranteed, i.e.\ worst case complexity, while the former refers
to the {\em typical} case, averaged over many instances of the TSP.

\begin{algorithm}[ht]
\caption{Procedure {\bf search} ($S$, $m$) -- search for the minimum energy configuration
$S_{\rm opt}$ within the subset ($S$, $m$) of all configurations.}
\label{search}
\begin{algorithmic}[1]
\STATE $n \leftarrow N-2m$; \COMMENT{number of free elements $s_i$}
\IF[$>2$ seq.\ in subset]{$n > 1$}
  \IF[bound]{$E_b(S, m) \geq E(S_{\rm opt})$}
	   \STATE return;
	\ELSE[branch]
	   \STATE {\bf search} ($S$, $m+1$);
		 \STATE $s_{m+1}\leftarrow -s_{m+1}$; {\bf search} ($S$, $m+1$);
		 \STATE $s_{N-m}\leftarrow -s_{N-m}$; {\bf search} ($S$, $m+1$);
		 \STATE $s_{m+1}\leftarrow -s_{m+1}$; {\bf search} ($S$, $m+1$);
	\ENDIF
\ELSIF[2 seq.\ in subset]{$n = 1$}
  \IF{$E(S) < E(S_{\rm opt})$} 
     \STATE $S_{\rm opt} \leftarrow S$; 
  \ENDIF
	\STATE $s_{m+1}\leftarrow -s_{m+1}$;
	\IF{$E(S) < E(S_{\rm opt})$} 
     \STATE $S_{\rm opt} \leftarrow S$; 
  \ENDIF
\ELSE[1 seq.\ in subset]
  \IF{$E(S) < E(S_{\rm opt})$} 
     \STATE $S_{\rm opt} \leftarrow S$; 
  \ENDIF
\ENDIF
\end{algorithmic}
\end{algorithm}
In accordance with our symmetry-classes, we specify a set of feasible solutions
by fixing the $m$ left- and $m$
rightmost elements of the sequence. The $N-2m$ center elements are not
specified, i.e.\ the set contains $2^{N-2m}$ feasible solutions.
Given a feasible set specified by the $m$ border elements, 4 smaller sets are
created by fixing the elements $s_{m+1}$ and $s_{N-m}$ to $\pm1$.
This is the branching rule. It is applied recursively until all elements have
been fixed. The energy of the resulting sequence is compared to the minimum
energy found so far. If it is lower, the sequence is kept as the potential
ground state. After all $c(m) 2^{N-2m}$ sequences have been considered, the
potential ground state has turned into a true one.

Lower bounds are usually obtained by replacing the original problem over a
given subset with an easier (relaxed) problem such that the solution value
of the latter bounds that of the former. A good relaxation is one that: (i) is
easy and fast to solve; and (ii) yields strong lower bounds. Most often
these are conflicting goals.

A relaxation of the LABS problem is
given by the problem to adjust the free elements (i.e.\ the center elements
$s_{m+1},\ldots,s_{N-m}$) to minimize all values $C_k^2$ {\em separately},
i.e.\ we replace the original problem
\begin{equation}
\label{hard_problem}
E_{{\rm min}} = \min_{{\rm subset}}\left(\sum_{k=1}^{N-1} C_k^2\right)
\end{equation}
with the relaxed version
\begin{equation}
\label{relaxation}
E_{{\rm min}}^* = \sum_{k=1}^{N-1} \min_{{\rm subset}}(C_k^2) \leq E_{{\rm min}}.
\end{equation}
Unfortunately $E_{{\rm min}}^*$ is still not easy to calculate, but we can proceed
our relaxation by providing a lower bound $E_b\leq E_{{\rm min}}^*$.
For that purpose we choose an arbitrary sequence from the subset with correlations $C_k$.
Complementing a free element $s_i\mapsto-s_i$ can decrement $|C_k|$ at most
by 2. Let
$f_k$ denote the number of terms $s_is_{i+k}$ in $C_k$ that contain at least
one free element. This leads to
\begin{eqnarray}
\label{lower_bound}
E_b &=& \sum_{k=1}^{N-k} \min\{b_k, (|C_k|-2f_k)^2\}\\
&\leq& E_{{\rm min}}^* \leq E_{{\rm min}}\nonumber 
\end{eqnarray}
where $b_k = (N-k)\bmod 2 \in\{0,1\}$ is the minimum value $|C_k|$ can attain.
The $f_k$ are given by
\begin{equation}
\label{free_terms}
f_k = \left\{
\begin{array}{cl}
0 & k \geq N-m \\
2(N-m-k) & N/2 \leq k < N-m\\
N-2m & k < N/2
\end{array}
\right.
\end{equation}
i.e.\ the long range correlations are not affected by our relaxation.
$E_b$ is not the strongest bound to $E_{\rm min}$,
but its calculation is very fast.

Now we have gathered all ingredients to formulate the branch and bound procedure 
{\bf search} (algorithm \ref{search}).
This procedure is called with two
parameters specifying the subset to search: a binary sequence
$S=\{s_1,\ldots,s_N\}$ and an integer $m$.
The subset consists of all $2^{N-2m}$
sequences that can be generated from $S$ by varying the $N-2m$ center
elements. $S_{\rm opt}$ is a global variable that holds the sequence with the minimum 
energy found so far. On entry, the size of the subset is checked: If it
contains more than 2 sequences, branch and bound (lines 3--10) is applied.
Otherwise the sequences in the subset are evaluated (lines 11--23).

The procedure {\bf search} is called from a driving procedure with $c(m_0)$
subsets, each representing a different symmetry-class. In practice, we used
$m_0 = 6$ with $528$ ($N$ even) resp.\ $544$ ($N$ odd) symmetry-classes.

\begin{figure}[htb]
  \includegraphics[width=\columnwidth]{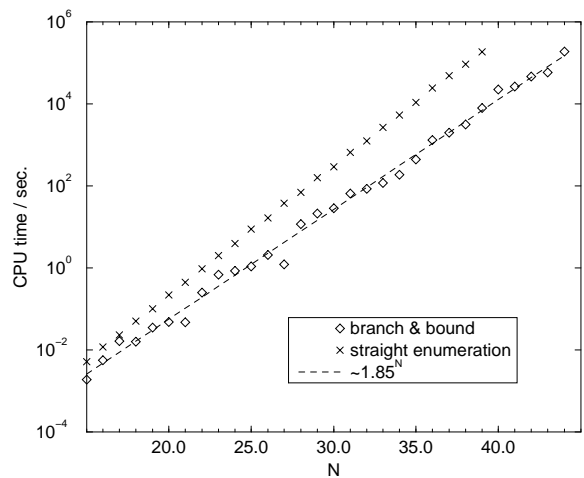}
 	\caption[Fig1]{\label{times}CPU time for exhaustive search algorithm vs.\ $N$.
	Times are measured on a Sun UltraSparc I 170 workstation.}
\end{figure}

To measure the impact of branch and bound, we started two runs on the
same machine: one ``straight'' enumeration (omitting lines 3--5) and the other
with activated bound-mechanism.
Figure \ref{times} displays the effect of branch and bound on the CPU time.
The straight enumeration shows the expected $O(2^N)$ behavior. Branch and bound reduces
the complexity to $O(1.85^N)$. Albeit this is still
exponential, the gain in speed is worth the little effort. The branch and bound
enumeration for $N=44$ took about 2 days on a Sun UltraSparc I
170 workstation. This compares well
with the extrapolated 68 days for the straight enumeration!

\subsection{Parallelization}

The different symmetry classes can be searched independently. Hence the straight
enumeration is perfectly parallelizable into $c(m)$ threads of control.
Branch and bound complicates the
situation: Whenever a better sequence is found by one thread, it
should be communicated immediately to all other threads to ensure that always
the best $E(S_{\rm opt})$ is used in the bounding test (line 3).
But $E(S_{\rm opt})$ is accessed very frequently, so the necessary
synchronization would spoil the parallelization. Giving each thread
its own local copy of $S_{\rm opt}$ preserves perfect parallelization but
abandons most of the benefits of branch and bound!

\begin{figure}[ht]
  \includegraphics[width=\columnwidth]{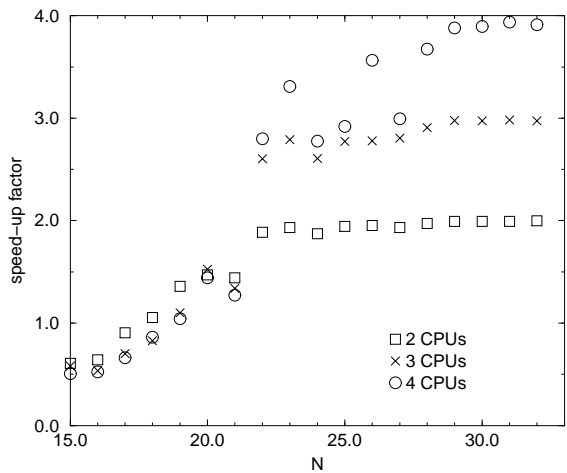}
 	\caption[Fig2]{\label{mt}Speed-up factor of the branch and
   bound algorithm on a symmetric multiprocessor platform using 2, 3 or 4 CPUs.}
\end{figure}

A solution to this dilemma is provided by the workpile paradigm
\cite{kleiman:shah:smaalders:96}: The symmetry classes to be searched for
are put on a central workpile and a number of worker threads are launched
together. Each worker thread requests an assignment of work from the workpile
(i.e.\ a symmetry class), performs the search, and then asks for a new
work assignment. This process repeats until all symmetry classes have been
considered.

The access to the workpile has to be protected with a mutual exclusion lock,
allowing only one thread at a time to read or modify data from the workpile.
If each worker thread uses its own local copy of $S_{\rm opt}$, this is the
only synchronization needed. To propagate the best $S_{\rm opt}$ as fast
as possible among the workers, it is stored in the workpile. A worker that
requests a new work assignment compares its own local $S_{\rm opt}$ with the global 
one and updates the one with the higher energy under the protection of the lock. 
This method limits the use of a suboptimal $S_{\rm opt}$ to the
search within $n-1$ symmetry classes, where $n$ is the number of worker
threads. The delay in propagation of the optimal $S_{\rm opt}$ is minimized
by choosing $c(m) \gg n$. In this case, the workpile paradigm has the
additional advantage of evenly distributing the load among all worker threads.
Due to branch and bound, the enumerations in some symmetry classes may take
considerably less time than in others. A worker that encounters these
``easy'' classes simply gets more classes to search.

On a 4 processor Sun SPARCstation 20, the number of worker threads ($\leq 4$)
is much smaller than $c(m)$ for $m = 6$, so the workpile paradigm should
yield almost perfect parallelization. Figure \ref{mt} shows that this is
indeed the case. The low speed-up factors for small $N$ are due to the
relative costs of thread-generation and synchronization compared to the actual
enumeration.

\section{Results}

Using the multithreaded branch and bound algorithm and 313 hours of CPU time on a
4-processor Sun SPARCstation 20, the ground
states of the Bernasconi-model have been found up to $N = 48$ (table
\ref{configs}). The enumeration for $N=32$ (the previous peak value) took
only 80 seconds, $N=39$ was done in 1 hour.
\begin{table}[p]
\begin{tabular}{rrl}
$N$ & $E_{\rm min}$ & sequence \\[1ex]
 3 &        1 & 21\\
  4 &        2 & 211\\
  5 &        2 & 311\\
  6 &        7 & 1113\\
  7 &        3 & 1123\\
  8 &        8 & 12113\\
  9 &       12 & 42111\\
 10 &       13 & 22114\\
 11 &        5 & 112133\\
 12 &       10 & 1221114\\
 13 &        6 & 5221111\\
 14 &       19 & 2221115\\
 15 &       15 & 52221111\\
 16 &       24 & 225111121\\
 17 &       32 & 252211121\\
 18 &       25 & 441112221\\
 19 &       29 & 4111142212\\
 20 &       26 & 5113112321\\
 21 &       26 & 27221111121\\
 22 &       39 & 51221111233\\
 23 &       47 & 212121111632\\
 24 &       36 & 2236111112121\\
 25 &       36 & 337111121221\\
 26 &       45 & 21212111116322\\
 27 &       37 & 34313131211211\\
 28 &       50 & 34313131211212\\
 29 &       62 & 212112131313431\\
 30 &       59 & 551212111113231\\
 31 &       67 & 7332212211112111\\
 32 &       64 & 71112111133221221\\
 33 &       64 & 742112111111122221\\
 34 &       65 & 842112111111122221\\
 35 &       73 & 7122122111121111332\\
 36 &       82 & 3632311131212111211\\
 37 &       86 & 844211211111122221\\
 38 &       87 & 8442112111111122221\\
 39 &       99 & 82121121234321111111\\
 40 &      108 & 44412112131121313131\\
 41 &      108 & 343111111222281211211\\
 42 &      101 & 313131341343112112112\\
 43 &      109 & 1132432111117212112213\\
 44 &      122 & 525313113111222111211121\\
 45 &      118 & 82121121231234321111111\\
 46 &      131 & 823431231211212211111111\\
 47 &      135 & 923431231211212211111111\\
 48 &      140 & 3111111832143212221121121
\end{tabular}
\caption[Tab1]{\label{configs}Ground states of the Bernasconi-model for
$3\leq N \leq 48$. Sequences are written in run-length notation: Each figure
indicates the number of consecutive equal signed elements.}
\end{table}
It is remarkable that from the 22 optimal skew-symmetric sequences in the
range $5 \leq N \leq 47$
\cite{golay:77}
7 have energies well above the true ground-state energie - a third.
This should be kept in mind if one uses skew-symmetric sequences to
estimate the groundstate energy in the limit $N\to\infty$.
\begin{figure}[ht]
  \includegraphics[width=\columnwidth]{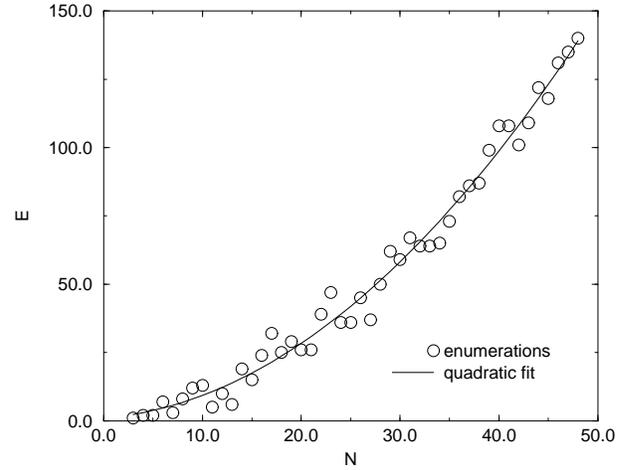}
 	\caption[Fig3]{\label{emin}Groundstate energy of the
Bernasconi-Hamiltonian vs.\ $N$.}
\end{figure}

Figure \ref{emin} shows the groundstate energies $E$ vs.\ $N$. In contrast
to the model with periodic boundary conditions
 there are no visible regular patterns for
special values of $N$ \cite{marinari:parisi:ritort:94a}. The energies seem to follow $E\propto N^2$ for all
values of $N$. A quadratic fit yields
\begin{equation}
F = \lim_{N\to\infty}\frac{N^2}{2E} = 9.3
\end{equation}
and leads us to the tentative conclusion that
\begin{equation}
F = \lim_{N\to\infty}\frac{N^2}{2E} > 9.
\end{equation}
This estimate is in agreement with Golay's conjecture $F\lesssim 12.32$ and
has to be compared to the value $F\approx 6.0$ found by heuristic searches
for long skew-symmetric sequences \cite{beenker:claasen:hermens:85} and by
simulated annealing \cite{bernasconi:87}. This indicates once more that 
heuristic and probabilistic methods fail to find the groundstates of the
Bernasconi-model. Every algorithm of such a kind should be jugded by the
percentage of values it finds from table \ref{configs}.

{\bf Acknowlegdements:} Thanks are due to A.~Engel and J.~Richter for guiding the
author's attention to the wonderful world of branch and bound and to 
S.~Kobe for providing helpful references.

\bibliographystyle{unsrt}
\bibliography{bib/labs,bib/math,bib/computer,bib/compphys}

%\end{multicols}
\end{document}